\documentclass[aip,pop]{revtex4-1}

\usepackage{graphicx,amsmath,amssymb}
\usepackage{dcolumn}
\usepackage{bm,color,setspace}

\begin{document}


\title{Particle-in-cell simulation study of the scaling of asymmetric
  magnetic reconnection with in-plane flow shear}

\author{C.~E.~Doss}
\author{P.~A.~Cassak}%
 \email{Paul.Cassak@mail.wvu.edu}
\affiliation{ 
Department of Physics and Astronomy, West Virginia University, Morgantown, 
West Virginia, 26506, USA}

\author{M.~Swisdak}
\affiliation{ 
Department of Physics and Institute for Research in Electronics and Applied Physics, University of Maryland, College Park, 
Maryland, 20742, USA
}%

\date{\today}

\begin{abstract}

We investigate magnetic reconnection in systems simultaneously
containing asymmetric (anti-parallel) magnetic fields, asymmetric
plasma densities and temperatures, and arbitrary in-plane bulk flow of
plasma in the upstream regions.  Such configurations are common in the
high-latitudes of Earth's magnetopause and in tokamaks.  We
investigate the convection speed of the X-line, the scaling of the
reconnection rate, and the condition for which the flow suppresses
reconnection as a function of upstream flow speeds.  We use
two-dimensional particle-in-cell simulations to capture the mixing of
plasma in the outflow regions better than is possible in fluid
modeling.  We perform simulations with asymmetric magnetic fields,
simulations with asymmetric densities, and simulations with
magnetopause-like parameters where both are asymmetric.  For flow
speeds below the predicted cutoff velocity, we find good scaling
agreement with the theory presented in Doss et al., {\it
  J.~Geophys.~Res.}, {\bf 120}, 7748 (2015).  Applications to
planetary magnetospheres, tokamaks, and the solar wind are discussed.
\end{abstract}

\maketitle


\section{\label{sec:intro}Introduction}

Magnetic reconnection is the fundamental plasma process where a change
in magnetic topology facilitates the conversion of magnetic energy to
plasma kinetic energy and heat.  It plays a fundamental role in
causing eruptions in the coronae of the sun and other stars, in the
interaction between the solar wind and the magnetospheres of Earth and
other planets, for confinement in toroidal fusion devices, and in a
large collection of astrophysical settings \cite{Zweibel09}.

There has been increased interest in the properties of reconnection in
realistic systems going beyond the simplifying assumptions of the
Sweet-Parker model \cite{Sweet58,Parker57}.  In this classical model,
the magnetic fields, densities, and temperatures are equal on either
side of the reconnection site, and the upstream plasmas has no bulk
flow other than the inflow in the reference frame of the reconnection
site.  One example of going beyond this model is to allow asymmetries
in the reconnecting magnetic fields, densities and temperatures on
either side of the reconnecting current sheet.  A second example is
including the effect of a bulk flow in the upstream plasma, whether in
the direction of the reconnecting magnetic field or out of the
reconnection plane.

Understanding how these effects impact the reconnection process, both
quantitatively and qualitatively, is often of great importance for
applying our understanding of reconnection to real systems.  One
example is reconnection at the dayside magnetopauses of Earth and
other planets.  The plasmas at the magnetopauses of Earth
\cite{Levy64} and Mercury \cite{DiBraccio13} differ on the two sides
and the solar wind drives a bulk flow in Earth's magnetosheath
\cite{Gosling86} and undoubtedly does at Mercury's, as well.  When the
interplanetary magnetic field (IMF) is northward, the magnetosheath
flow is parallel/anti-parallel to the reconnecting magnetic field in
the polar regions.  When the IMF is southward, magnetosheath flow at
the flanks is mostly out of the reconnection plane.  The effect of
upstream bulk flow is even more dramatic at the magnetospheres of
Jupiter and Saturn, where rotation of the magnetosphere is much
stronger of an effect than at Earth \cite{Vasyliunas83}.

A second example where upstream asymmetries and bulk flow are
important is in tokamaks.  The density and temperature profiles are
peaked in the plasma core with a spatially varying magnetic field, so
the plasma profiles at the reconnection site are non-uniform
\cite{Mirnov06}.  Further, there are often bulk flows causing the
toroidal and poloidal rotation of the plasma \cite{LaHaye10},
especially those driven by neutral beam injection.  Therefore, both
asymmetries and flows effects are present and are important to the
dynamics in magnetospheres and tokamaks.

While the effect of asymmetries and flow shear have separately
received much attention \cite{Cassak16}, only a few studies have
treated systems that simultaneously contain both effects.  Studies of
the shock structure far downstream of the reconnection site were
carried out analytically \cite{Heyn88,Biernat89,Lin94} and using
magnetohydrodynamic (MHD) modeling \cite{LaBelleHamer95}.
Particle-in-cell simulations were used to study systems simultaneously
including asymmetries, flow shear, and an out-of-plane (guide)
magnetic field \cite{Tanaka10}.  It was shown that the flow shear and
diamagnetic drifts set up with the pressure gradient and the guide
field can either reinforce or counteract each other.

More recently, a scaling analysis for systems including both
asymmetries and upstream flow in the reconnection plane was performed
\cite{Doss15}.  It was argued that the reconnection site (the X-line)
typically convects in the outflow direction.  The convection speed of
the X-line and the rate of reconnection was predicted as a function of
arbitrary upstream plasma parameters for isolated systems; the results
will be reviewed in Sec.~\ref{sec:theory}.  In symmetric reconnection
with a flow shear, reconnection does not occur if the flow is
super-Alfv\'enic because the tension in the reconnecting magnetic
field cannot overcome the energy of the flow \cite{LaBelleHamer94}.
There is also a critical flow speed above which reconnection does not
occur for asymmetric reconnection; a generalization of the symmetric
result for the asymmetric case was also derived \cite{Doss15}.  These
predictions were successfully tested with two-dimensional numerical
simulations using the two-fluid model (MHD with the Hall term and
electron inertia).  However, it is known that the fluid model is not
well-suited to describe systems with asymmetric density and
temperature as the fluids do not mix in the absence of thermal
conduction \cite{Cassak09c,Birn10,Ouellette14}; even if conduction is
present, the fluid model may not faithfully describe mixing in a
nearly collisionless plasma as is the case in many applications.
These shortcomings are not present in kinetic simulations, such as the
particle-in-cell numerical technique \cite{Birdsall91a} where
macro-particles are evolved in time and plasma mixing naturally
occurs.  Thus, it is important to investigate the scaling of the
reconnection rate and the drift speed of isolated X-lines within a
fully kinetic model.

In this study, we perform a systematic numerical study of magnetic
reconnection with asymmetries and in-plane upstream flow using the
particle-in-cell (PIC) technique.  We measure relevant quantities in
independent simulations in which all quantities are held fixed other
than the upstream flow.  We find that the theoretical predictions
previously tested in fluid simulations \cite{Doss15} are consistent
with the results of the PIC simulations.

In Sec.~\ref{sec:theory}, we review the predictions for the convection
speed of isolated X-lines and the reconnection rate in terms of
upstream parameters.  Sec.~\ref{sec:simulations} discusses the
simulations we perform as well as our methodology for analyzing the
simulation data.  Sec.~\ref{sec:results} presents the simulation
results and compares them to the predictions.  We summarize our
results and discuss applications of the results in
Sec.~\ref{sec:discussion}.

\section{\label{sec:theory}Theory}

Scaling laws for the dissipation region's convection speed and the
reconnection rate were derived \cite{Doss15} for isolated
configurations including asymmetric (but anti-parallel) magnetic
fields and asymmetric densities and temperatures, along with arbitrary
in-plane upstream flow.  We define the magnetic field strengths of the
two upstream regions as $B_{L,1}$ and $B_{L,2}$, the plasma mass
densities as $\rho_1$ and $\rho_2$, and the upstream flow speeds as
$v_{L,1}$ and $v_{L,2}$, where the $L$ subscript is borrowed from
boundary normal coordinates in magnetospheric applications to denote
the reconnecting component, and the 1 and 2 subscripts denote the two
upstream sides of the reconnection site.  The reconnecting magnetic
fields are treated as positive quantities, and the speeds are defined
as positive if in the direction of $B_{L,1}$ and negative in the
direction of $B_{L,2}$.

The convection speed $v_{{\rm drift}}$ of the X-line along the current
sheet, the reconnection rate $E_{{\rm shear,asym}}$, and the upstream
flow speed $v_{{\rm shear,crit}}$ required for steady state
reconnection to be prevented scale (in cgs units) as
\begin{equation}
  v_{{\rm drift}}\sim\frac{\rho_1B_{L,2}v_{L,1}+\rho_2B_{L,1}v_{L,2}}
  {\rho_1B_{L,2}+\rho_2B_{L,1}},
\label{eqn::drift_eqn}
\end{equation}
\begin{equation}
  E_{{\rm shear,asym}}\sim E_{{\rm asym,0}} \left(1-\frac{v_{{\rm
        shear}}^2} {c_{A,{\rm asym}}^2} \frac{4 \rho_1 B_{L,2} \rho_2
    B_{L,1}} {(\rho_1 B_{L,2} + \rho_2 B_{L,1})^2} \right),
\label{eqn::recon_eqn}
\end{equation}
and
\begin{equation}
  v_{{\rm shear,crit}} \sim c_{A,{\rm asym}} \frac{\rho_1 B_{L,2} + \rho_2
    B_{L,1}} {2 (\rho_1 B_{L,2} \rho_2 B_{L,1})^{1/2}}.
\label{eqn::vcrit}
\end{equation}
In writing these expressions, the asymmetric Alfv\'en speed $c_{A,{\rm
    asym}}$, the asymmetric reconnection rate in the absence of a flow
shear $E_{{\rm asym},0}$, and the velocity shear $v_{{\rm shear}}$ are
\begin{equation}
  c_{A,{\rm
      asym}}^2\sim\frac{B_{L,1}B_{L,2}}{4\pi}\frac{B_{L,1}+B_{L,2}}{\rho_1B_{L,2}
    + \rho_2B_{L,1}},
\label{eqn::alfven_outflow}
\end{equation}
\begin{equation}
  E_{{\rm asym,0}} = \frac{B_{L,1} B_{L,2}}{B_{L,1} + B_{L,2}}
  \frac{c_{A,{\rm asym}}}{c}\frac{2 \delta}{L_d},
\label{eqn::asym_rate}
\end{equation}
and
\begin{equation}
  v_{{\rm shear}} = \frac{v_{L,1} - v_{L,2}}{2},
\label{eqn::shear_vel}
\end{equation}
where $\delta$ and $L_d$ are the half-thickness in the normal
direction and half-length in the outflow direction of the dissipation
region and $c$ is the speed of light.

Equation~(\ref{eqn::drift_eqn}) was derived using conservation of
momentum in the $L$ direction into and out of the dissipation region.
Equation~(\ref{eqn::recon_eqn}) follows from treating the energetics
of the release of magnetic tension on the outflow jets while including
the upstream flow, keeping track of the fact that the X-line and
stagnation point are not colocated in asymmetric reconnection
\cite{Cassak07d}.  Equation~(\ref{eqn::vcrit}) is the condition that
makes $E_{{\rm shear,asym}} = 0$ in Eq.~(\ref{eqn::recon_eqn}), which
is the condition for when reconnection shuts off.
Equations~(\ref{eqn::alfven_outflow}) and (\ref{eqn::asym_rate})
follow from the analysis of asymmetric reconnection with no upstream
flow \cite{Cassak07d}.  Equation~(\ref{eqn::shear_vel}) is merely
shorthand for the quantity of import for the reconnection rate in
Eq.~(\ref{eqn::recon_eqn}).

We emphasize a few important assumptions made in this analysis.  It
assumed no upstream out-of-plane (guide) magnetic field.  It also
assumed that the outflow speeds in the two downstream directions are
equal and opposite in the reference frame of the moving X-line.  As a
scaling analysis, it assumes a single characteristic value represents
each quantity in the appropriate regions.

\section{\label{sec:simulations}Simulations}

We perform 2D kinetic particle-in-cell simulations using the P3D code
\cite{Zeiler02} to test the predictions.  Particles are stepped
forward using a relativistic Boris algorithm, while electromagnetic
fields are updated with a second order trapezoidal leapfrog.  Magnetic
field strengths are normalized to an arbitrary strength $B_0$ and
plasma number densities are normalized to an arbitrary density $n_0$.
Values of length and speed are normalized to the ion inertial length
$d_{i0}=(m_i c^{2} / 4 \pi n_0 e^{2})^{1/2}$ and the Alfv\'en speed
$c_{A0} = B_0 / (4 \pi m_i n_0)^{1/2},$ respectively, where $e$ is the
ion charge and $m_i$ is the ion mass.  The unit of time is therefore
$t_0=d_{i0}/c_{A0}=\Omega_{ci0}^{-1}$.

As in the fluid simulations \cite{Doss15}, the boundary conditions are
doubly periodic, and the magnetic field profile is initialized as a
double Harris sheet:
\begin{equation}
  B_{x}(y) = \left\{ \begin{array}{ll} -B_{L,1}
    \tanh\left(\frac{|y|-L_{y}/4}{w_{0}}\right) & \hskip 0.1cm L_{y} /
    4 < |y| < L_{y} /2 \\ -B_{L,2}
    \tanh\left(\frac{|y|-L_{y}/4}{w_{0}}\right) & \hskip 0.52cm 0 <
    |y| < L_{y}/4 \end{array} \right.
\label{eqn::B_init}
\end{equation}
where $w_0 = 1.0 \ d_{i0}$ is the initial current sheet width and
$L_y$ is the domain size in the inflow direction.  (The $x$ direction
in the simulations corresponds to the $L$ direction in boundary normal
coordinates.)  There is no out-of-plane guide field.  The temperature
profile of species $j$, which can denote electrons $e$ or ions $i$, is
initialized as
\begin{equation}
  T_j(y) = \frac{T_{j1} + T_{j2}}{2} + \frac{T_{j1} - T_{j2}}{2}
  \tanh\left(\frac{|y|-L_{y}/4}{w_{0}}\right),
\label{eqn::rho_init}
\end{equation}
where $T_{j1}$ and $T_{j2}$ are selected asymptotic initial
temperatures.  We use $T_{i1} / T_{e1} = T_{i2} / T_{e2} = 2$ for all
simulations.  Initial electron and ion densities are chosen to be
equal with asymptotic values of $n_{01}$ and $n_{02}$.  The density
profiles initially enforce pressure balance across the current sheet
in the fluid sense.  There is no known general asymmetric kinetic
equilibrium \cite{Swisdak03,Pritchett08} (although there are
approximations \cite{Belmont12} that are not employed here).  As in
many previous studies, our system rings at early times but settles to
a steady state by the time of interest for this study.  By this time,
the initial kinks in the initial magnetic field and bulk velocity
profiles at the current sheets also smooth out and therefore do not
present any problems.

\begin{table}
\caption{Initial upstream plasma parameters for the simulations in
  this study.  The set labeled ``$B$'' have asymmetric fields with a
  symmetric density.  The set labeled ``$n$'' has a symmetric magnetic
  field and an asymmetric density.  The set labeled ``$ms$'' is
  representative of Earth's magnetopause.  The predicted critical flow
  shear to shut off reconnection is given for each set.}
\begin{tabular}{c | c c c c c c c c | c}
  	\hline
  	$Set$ & $B_{L,1}$ & $B_{L,2}$ & $n_{01}$ & $n_{02}$ & $T_{e1}$ & $T_{i1}$ & $T_{e2}$ & $T_{i2}$ & $v_{{\rm shear,crit}}$ \\
 	\hline
	$B$  & 1.5 & 0.5 & 0.2 & 0.2 & 0.667 & 1.333 & 2.333 & 4.667 & 2.2\\
	$n$  & 1.0 & 1.0 & 0.6 & 0.2 & 0.667 & 1.333 & 2.000 & 4.000 & 1.8 \\
	$ms$ & 1.0 & 2.0 & 1.0 & 0.1 & 0.667 & 1.333 & 1.667 & 3.333 & 4.0 \\
        \hline
\end{tabular}
\label{table::runparms}
\end{table}

The ion and electron bulk flow speeds are initialized as a double
tanh profile:
\begin{equation}
  v_{j,x}(y) = \left\{ \begin{array}{ll} -v_{L,1}
    \tanh\left(\frac{|y|-L_{y}/4}{w_{0}}\right) & \hskip 0.1cm L_{y} /
    4 < |y| < L_{y} /2 \\ -v_{L,2}
    \tanh\left(\frac{|y|-L_{y}/4}{w_{0}}\right) & \hskip 0.52cm 0 <
    |y| < L_{y}/4.  \end{array} \right.
\label{eqn::shear_init}
\end{equation}
and there is no out-of-plane component of the flow.  This is
accomplished by loading particles with a Harris-type drifting
Maxwellian distribution function with a non-zero $v_{j,x}$
contribution given by Eq.~(\ref{eqn::shear_init}), which is equivalent
to a previously used approach \cite{Roytershteyn08}.  The electron and
ion bulk flow speeds in the $x$ direction are identical.

\begin{table}
\caption{Simulations performed in this study, relevant predicted
  quantities, and measured quantities from the simulations.  $v_{L,1}$
  and $v_{L,2}$ are initial upstream flow speeds for the three sets of
  simulations discussed in Table~\ref{table::runparms}.  Measured
  values of $v_{{\rm drift}}$ and $E$ from the simulations are given
  for the top (T) and bottom (B) current sheets.  Entries with blank
  values did not reconnect in the standard way.  Predictions for the
  X-line convection speed and reconnection rates from
  Eqs.~(\ref{eqn::drift_eqn}) and (\ref{eqn::recon_eqn}) are labeled
  $v_{{\rm drift,pred}}$ and $E_{{\rm pred}}$, respectively.  The
  value for $E_{{\rm asym},0}$ is taken from the averaged measured
  values for the case with no upstream flow, {\it i.e.,} $(E_T +
  E_B)/2$.}
\begin{tabular}{ c c c | c c c | c c c }
  	\hline
  	$Set$ & $v_{L,1}$ & $v_{L,2}$ & $v_{{\rm drift,pred}}$ & $v_{{\rm drift},T}$ & $v_{{\rm drift},B}$ & $E_{{\rm pred}}$ & $E_T$ & $E_B$ \\
 	\hline
	$B$ & 0.0 & 0.0 & 0.0 & -0.086 & -0.082 & 0.060 & 0.054 & 0.065 \\
	$B$ & 0.2 & -0.2 & 0.1 & -0.026 & 0.11 & 0.059 & 0.056 & 0.055 \\
	$B$ & 0.4 & -0.4 & 0.2 & 0.12 & 0.15 & 0.058 & 0.061 & 0.052 \\
	$B$ & 0.6 & -0.6 & 0.3 & 0.17 & 0.19 & 0.055 & 0.058 & 0.058 \\
	$B$ & 0.8 & -0.8 & 0.4 & 0.30 & 0.35 & 0.052 & 0.051 & 0.057 \\
	$B$ & 1.2 & -1.2 & 0.6 & 0.55 & 0.52 & 0.042 & 0.038 & 0.036 \\
	$B$ & 1.6 & -1.6 & 0.8 & 0.68 & 0.67 & 0.029 & 0.025 & 0.025 \\
	$B$ & 2.0 & -2.0 & 1.0 & 0.85 & 0.85 & 0.012 & 0.017 & 0.021 \\
	$B$ & 2.4 & -2.4 &  &  &  &  &  &  \\
	$B$ & 2.8 & -2.8 &  &  &  &  &  &  \\
	$B$ & 2.0 & 2.0 & 2.0 & 1.83 & 1.96 & 0.060 & 0.057 & 0.062 \\
	\hline
	$n$ & 0.0 & 0.0 & 0.0 & 0.033 & -0.015 & 0.099 & 0.097 & 0.10 \\
	$n$ & 0.4 & -0.4 & 0.2 & 0.11 & 0.11 & 0.94 & 0.095 & 0.095 \\
	$n$ & 0.8 & -0.8 & 0.4 & 0.27 & 0.30 & 0.080 & 0.075 & 0.080 \\
	$n$ & 1.2 & -1.2 & 0.6 & 0.50 & 0.48 & 0.056 & 0.054 & 0.050 \\
	$n$ & 1.6 & -1.6 & 0.8 & 0.67 & 0.84 & 0.023 & 0.037 & 0.040 \\
	$n$ & 2.0 & -2.0 &  &  &  &  &  &  \\
	$n$ & 2.4 & -2.4 &  &  &  &  &  &  \\
	\hline
	$ms$ & 1.0 & 0.0 & 0.95 & 0.85 & 0.90 & --- & 0.14 & 0.14 \\
	$ms$ & 2.0 & 0.0 & 1.90 & 1.75 & 1.53 & --- & 0.12 & 0.13 \\
	\hline
\end{tabular}
\label{table::runs}
\end{table}

For all simulations, the speed of light is $c=15 \ c_{A0}$ and the
electron mass is $m_e=m_i/25$.  The time step for particles is
$dt=0.006 \ \Omega_{ci0}^{-1}$ and the electromagnetic fields have a
time step half as much.  The grid scale is $dx = dy = 0.05 \ d_{i0}$.
The simulations performed for this study are summarized in
Table~\ref{table::runparms}.  The set labeled ``$B$'' employ
asymmetric magnetic fields with symmetric density, the set labeled
``$n$'' have symmetric magnetic fields with asymmetric density, and
the set labeled ``$ms$'' are for representative magnetospheric
conditions \cite{Malakit10}.  The domain size is $L_x \times L_y =
204.8 \times 102.4 \ d_{i0}$ for the $B$ simulations and $L_x \times
L_y = 102.4 \times 51.2 \ d_{i0}$ for the $n$ and $ms$ simulations.
The initial number of particles-per-grid cell is $1000$ for the $B$
and $n$ simulations and $500$ for the $ms$ simulations.  The upstream
flow speeds are varied for each set; Table~\ref{table::runs} shows the
values used in the present study.

To reach the nonlinear phase of reconnection more rapidly, the
simulations are initialized using a coherent divergence-free
sinusoidal perturbation to the magnetic fields of amplitude 0.1 with
one full wavelength of the perturbation in the $x$ direction and two
full wavelengths in the $y$ direction.  Each simulation is evolved
until magnetic reconnection reaches a steady-state.  Since the
reconnection rate differs for different upstream flow speeds, the
steady state is reached at different times for different simulations.
Consequently, we use the half-width of the primary magnetic island as
a common indicator across the simulations; a range of island widths of
4-6 $d_{i0}$ is used to identify comparable times.  If needed, the
interval is slightly adjusted to ensure the system is in a steady
state.

For each time step and each current sheet, the X-line and O-line are
found using standard techniques by calculating the flux function
$\psi$ as ${\bf B} = {\bf \hat{z}} \times \nabla \psi$.  The saddle
point of $\psi$ is the X-point and the extremum is the O-point.  The
convection velocity of the reconnection site is measured as the time
derivative of the X-line position.  The reconnection rate is the time
rate of change of the magnetic flux difference between the X-line and
O-line.  These values are averaged over the steady state interval to
provide a representative value for that simulation.  

\section{\label{sec:results}Results}

\begin{figure}
\includegraphics[width=20pc]{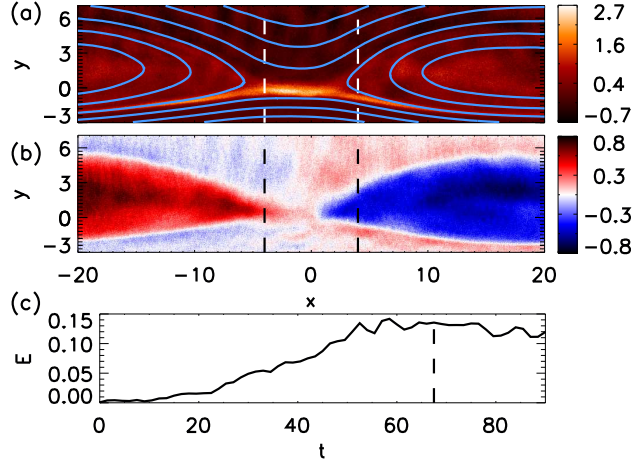}
\caption{\label{fig::panel_c} 2D plots of data from the
  magnetopause-like simulation with $v_{L,1} = 1$ and $v_{L,2} = 0$.
  (a) Out-of-plane current density $J_z$, with magnetic field lines in
  blue and (b) out-of-plane magnetic field $B_z$.  Plots show only a
  small portion of the computational domain. (c) Reconnection rate $E$
  as a function of time $t$ for this simulation.}
\end{figure}

We begin by showing an overview of the plasma parameters in a
simulation with representative magnetospheric conditions, specifically
the $ms$ simulation with $v_{L,1} = 1$ and $v_{L,2} = 0$.
Figure~\ref{fig::panel_c} contains (a) the out-of-plane current
density $J_z$ with magnetic field lines overplotted and (b) the out of
plane magnetic field $B_z$ at a time of 67.5 when the reconnection
rate has reached a steady state, with the coordinate system shifted so
that the X-line is at the origin.  Only a fraction of the total
computational domain is plotted.  Interestingly, the results are quite
similar to standard systems without flow shear.  In particular, the
Hall magnetic field in (b) is mostly bipolar, as is the norm in
strongly asymmetric systems \cite{Karimabadi99,Pritchett08}.  The
normal electric field $E_y$ (not shown) shows the typical asymmetric
Hall electric field dominated by a positive $E_y$ on the strong
(magnetospheric) field side of the dissipation region and a negative
Larmor electric field upstream of it \cite{Malakit13}.  Panel (c)
shows the reconnection rate $E$ as a function of time from this
simulation, showing that the system reaches a steady-state by
approximately $t = 60$.

\begin{figure}
\includegraphics[width=20pc]{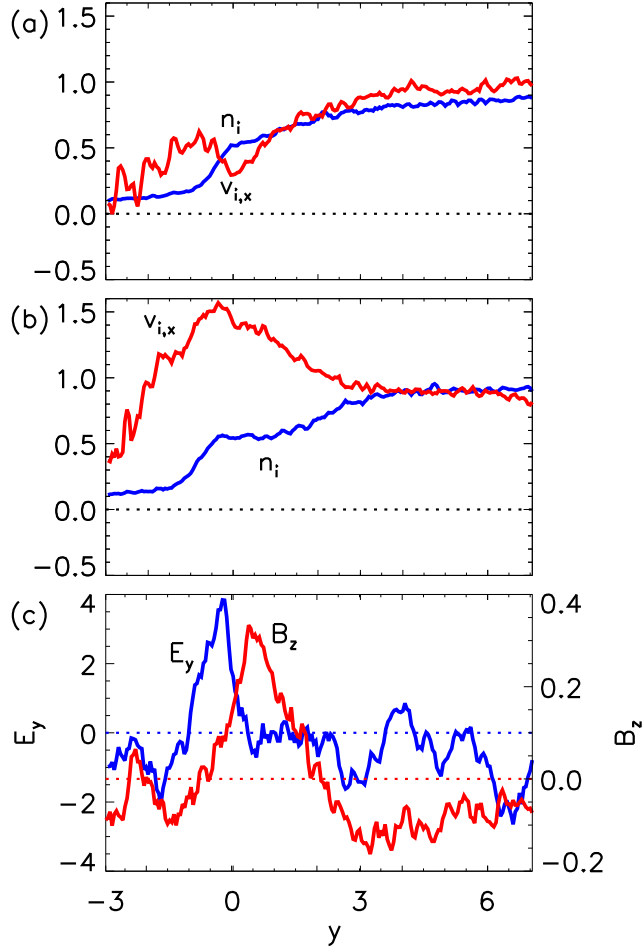}
\caption{\label{fig::panel_f} Cuts in the inflow ($y$) direction of
  plasma parameters from the simulation shown in
  Fig.~\ref{fig::panel_c}.  Plotted are the ion density $n_i$ and ion
  bulk flow $v_{i,x}$ in the direction of the reconnecting field $4
  d_{i0}$ to the (a) left and (b) right of the X-line.  (c) Normal
  electric field $E_y$ (blue) and out-of-plane magnetic field $B_z$
  (red) $4 d_{i0}$ to the left of the X-line.}
\end{figure}

Figure~\ref{fig::panel_f} shows cuts 4 $d_{i0}$ downstream of the
X-line from the same simulation.  The ion density $n_i$ and ion bulk
flow velocity $v_{i,x}$ to the left and right of the X-line are shown
in panels (a) and (b), respectively.  They reveal the negative and
positive deflections from the background flow profile due to the
reconnection exhausts.  Panel (c) shows $E_y$ on the left axis and
$B_z$ on the right axis in a cut 4 $d_{i0}$ to the left of the X-line.
The dashed horizontal lines mark zero for the two plots.  The Hall
magnetic ($y = 1$), Hall electric ($y = -0.5$) and Larmor electric
($y= -2$) fields are present.  These results suggest that the upstream
bulk flow largely does not alter the kinetic signatures of
reconnection for typical magnetospheric conditions.

Another feature of reconnection with an upstream flow shear is the
tilting of the current sheet near the X-line
\cite{LaBelleHamer94,LaBelleHamer95,Cassak11b}.  We see the current
sheet tilt in the present simulations, as well.  The tilting in the
$n$ simulations with asymmetric density is more pronounced than
similar upstream flows from the $B$ simulations with asymmetric
magnetic field (not shown).  The tilt is more pronounced for higher
flow shear, as is to be expected.

\begin{figure}
\noindent\includegraphics[width=20pc]{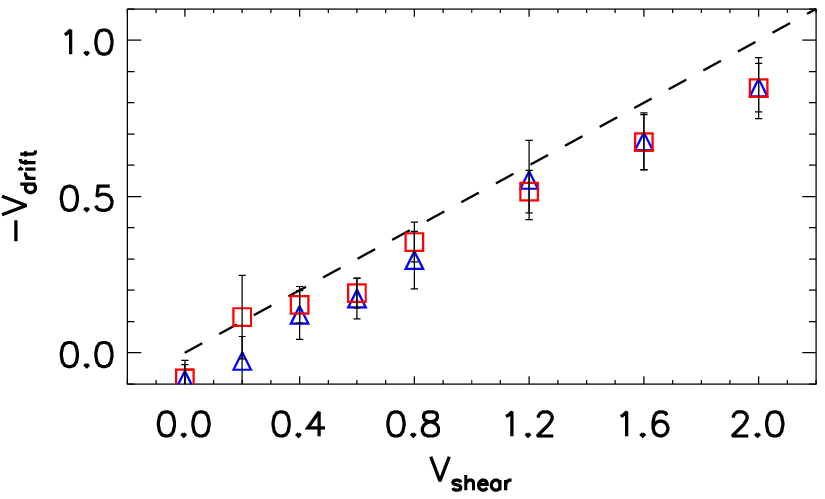}
\noindent\includegraphics[width=20pc]{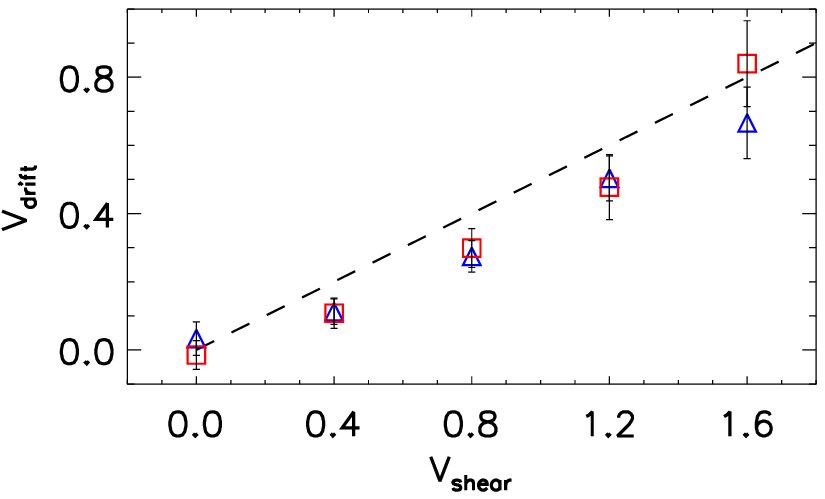}
\caption{\label{fig::magdrift} Convection speed $v_{{\rm drift}}$ of
  the X-line as a function of flow speed $v_{{\rm shear}}$ for
  simulations with (top) asymmetric magnetic field strengths $B_1=1.5$
  and $B_2=0.5$ and symmetric density 0.2 and (bottom) asymmetric
  densities $n_1=0.6$ and $n_2=0.2$ and symmetric magnetic field
  strength $B=1.0$.  The speeds in the top plot are negative because
  the X-line convects to the left.  Triangles and squares are for the
  top and bottom current sheets, respectively.  The predicted
  convection speed from Eq.~(\ref{eqn::drift_eqn}) is given by the
  dashed line.}
\end{figure}

Next, we test the predictions for the X-line drift speed, reconnection
rate, and cutoff speed.  Raw measured and predicted values for all
simulations are given in Table~\ref{table::runs}.  We first consider
the $B$ simulations containing asymmetries in magnetic field strength.
From Eq.~(\ref{eqn::vcrit}), the predicted cutoff speed is about $2.2
\ c_{A0}$.  We vary the upstream flow speed up to this cessation
condition, with $v_{L,1} = - v_{L,2}$ for simplicity, and measure the
drift speed in each simulation.  The results are shown in the top plot
in Fig.~\ref{fig::magdrift}.  Here and throughout, blue triangles and
red squares are for the two current sheets in the equilibrium.  Error
bars are determined as the standard deviation during the steady state
intervals.  A linear trend in drift speed for increasing shear flow
speed is observed, qualitatively consistent with
Eq.~(\ref{eqn::drift_eqn}).  The dashed black line is the prediction
of Eq.~(\ref{eqn::drift_eqn}), so the quantitative agreement is good
as well.

We carry out the same analysis on the asymmetric density simulations,
with results shown in the bottom plot of Fig.~\ref{fig::magdrift}.
The predicted cessation condition for these simulations is
approximately $1.8 \ c_{A0}$.  Again, the trend and absolute agreement
is quite good.  Note that some of the measurements from the simulation
are slightly below the prediction for both the $B$ and $n$
simulations; this could be due to the inertia of moving larger primary
islands in this finite sized and periodic domain.

\begin{figure}
\noindent\includegraphics[width=20pc]{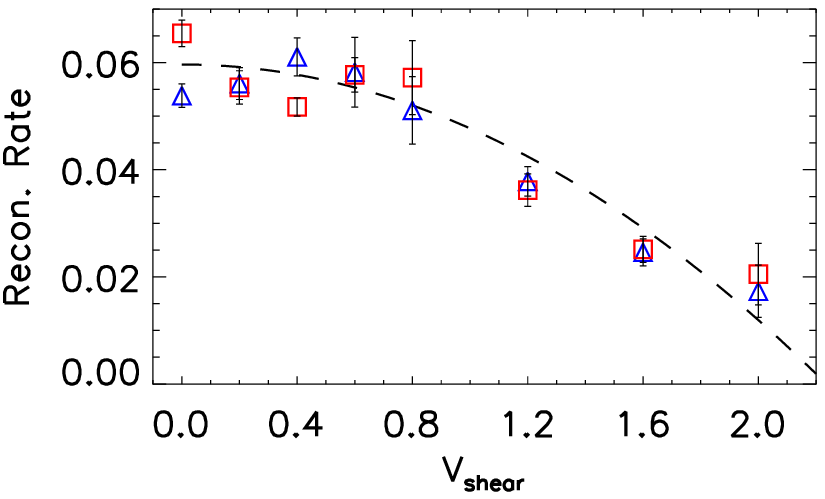}
\noindent\includegraphics[width=20pc]{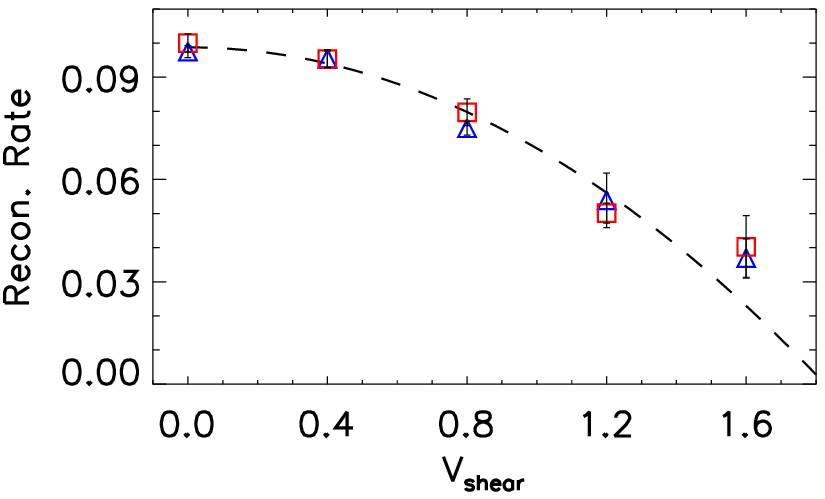}
\caption{\label{fig::magrate} Reconnection rate $E_{{\rm asym,shear}}$
  as a function of flow speed $v_{{\rm shear}}$ for simulations with
  (top) asymmetric magnetic field strengths $B_1=1.5$ and $B_2=0.5$
  and symmetric density $n=0.2$ and (bottom) asymmetric densities
  $n_1=0.6$ and $n_2=0.2$ and symmetric magnetic field strength
  $B=1.0$.  Triangles and squares are for the top and bottom current
  sheets, respectively.  The predicted reconnection rate from
  Eq.~(\ref{eqn::recon_eqn}) is given by the dashed line and is
  normalized to the average reconnection rate in the simulation with
  $v_{{\rm shear}} = 0$.}
\end{figure}

Next, we test the reconnection rates obtained in both sets of
simulations.  The results are plotted in Fig.~\ref{fig::magrate} for
(top) asymmetric magnetic field and (bottom) asymmetric density sets,
respectively.  The dashed lines denote the predicted values; in
calculating the predictions, we use the average reconnection rate from
the two sheets in the zero flow shear case as $E_{{\rm asym},0}$ in
Eq.~(\ref{eqn::recon_eqn}).  We find excellent agreement between these
measurements and the predictions.

A similar analysis is carried out for the representative
magnetospheric simulations.  From the prediction, the convection speed
should double and the reconnection rate should drop by $5\%$ as
$v_{L,1}$ is increased from $1.0$ to $2.0$.  The measured convection
speed is 0.88 for $v_{L,1} = 1.0$ and 1.64 for $v_{L,2} = 2.0$, which
differ by a factor of $1.86$; this agrees well with the prediction.
The reconnection rate drops from 0.143 to 0.123, a decrease of $13\%$.
While ostensibly greater than the prediction, the uncertainty in these
measurements is enough where the difference is not expected to be
significant.  The key is that the significant increase in upstream
flow speed only slightly impacts the reconnection rate, as predicted
for reconnection in systems with a strongly asymmetric density
\cite{Doss15}.

As a test of the prediction for the cutoff flow shear speed to shut
off asymmetric reconnection, we report results from simulations beyond
the predicted cessation conditions.  We run simulations with $v_{{\rm
    shear}}$ = 2.4 and 2.8 for the asymmetric field and $v_{{\rm
    shear}}$ = 2.0 and 2.4 for the asymmetric density simulations.
The current sheets in these systems tend to contort and become
sinusoidal rather than flat, as if beginning to go Kelvin-Helmholtz
unstable.  This is qualitatively different than the simulations below
the threshold, where reconnection clearly starts from the beginning.
We also point out that some of them nonlinearly experience
reconnection in the strongly bent fields.  We argue this form of
reconnection is different than the robust form of reconnection
discussed earlier, so we assert that the simulation results are
consistent with Eq.~(\ref{eqn::vcrit}).

Finally, we comment on whether the predictions remain valid when the
upstream flow on the two sides is in the same direction, {\it i.e.,}
both $v_{L,1}$ and $v_{L,2}$ are positive.  We do an asymmetric
magnetic field simulation with $v_{L,1} = v_{L,2} = 2$.  The
predictions are that the X-line will drift with the common upstream
drift speed and the reconnection rate is the same as if there was no
flow.  As shown in Table~\ref{table::runs}, this is borne out in the
simulations.  In summary, the simulation results agree quite well with
the predictions discussed in Sec.~\ref{sec:theory}.

\section{\label{sec:discussion}Discussion}

We use particle-in-cell simulations to study the scaling of 2D
asymmetric anti-parallel reconnection with an in-plane upstream flow.
The particle-in-cell approach is necessary to faithfully capture the
effect of plasma mixing between the two disparate plasmas in the
exhaust region, which is not well-described in the fluid description.
We find very good agreement with the scaling predictions as a function
of upstream plasma parameters for the drift speed of the X-line, the
reconnection rate, and the critical upstream flow speed necessary to
suppress reconnection found in a recent study \cite{Doss15}.

One area in which this study goes beyond the previous fluid simulation
study is by testing the theory for systems in which the flow is in the
same direction on each upstream side, which has been studied for solar
wind applications \cite{Einaudi01,Bettarini06,Bettarini09}.  The
results confirm the theory \cite{Doss15} works in this case as well,
and in particular confirms that the figure of merit is $v_{{\rm
    shear}} \propto v_{L,1} - v_{L,2}$.  Physically, for a case with
$v_{L,1}$ and $v_{L,2}$ in the same direction, the upstream plasma on
both sides enter the diffusion region with momentum in the direction
of the reconnecting field, and this momentum makes the X-line convect
at the weighted average of the two speeds.  In the special case of
equal flows on the two sides, the upstream plasmas are stationary in
the reference frame of the X-line, so there is no effect on the
reconnection rate.  This reveals that reconnection in the solar wind
should not be suppressed by flow shear, which is consistent with the
observation of active reconnection in the solar wind
\cite{gosling05a,gosling07a,Gosling12,Mistry15}.

A important consideration before applying the results here to
naturally occurring reconnection or reconnection in the laboratory is
that the X-lines in question must be ``isolated'' in the sense that
they are free to convect in the external flow for the theory to apply.
This is essentially satisfied in the solar wind, for example, and may
also be the case for neoclassical tearing modes (NTMs) in tokamaks.
However, for the magnetopause of Earth and other planets, one should
proceed with caution.  A primary X-line is undoubtedly controlled by
global considerations such as being line-tied to the ionosphere.
Therefore, it is not clear if a single X-line would follow the
predictions of the theory tested here.  However, a flux rope or flux
transfer event (FTE) \cite{Russell78} could be considered isolated, so
the theory may apply.  There are differences between the predictions
tested here and the leading model of open flux ({\it i.e.}, FTE)
motion \cite{cowley89a}, so future work on this topic would be
interesting.

An area of potential interest for future study are the properties of
the system in regimes where the flow is higher than the cessation
condition.  In our simulations, the current sheet contorts as if
beginning to undergo Kelvin-Helmholtz instability.  Then, reconnection
as a secondary process on the bent current sheets begins.  The current
sheets flatten and reconnect robustly.  It is not clear whether this
is physical or only a function of the finite system size in the
simulations, so it is worth future study.  We point out that a limit
with such strong flows is not likely to apply at the magnetosphere
except possibly when the reconnection site interacts with the dense,
cold plasmas in plasmaspheric drainage plumes
\cite{Borovsky06b,Walsh14,Walsh14b}, but may be important in tokamaks.

One limitation of this study is that it is in 2D.  Any 3D dynamics,
including drift waves set up with wave vector in the out-of-plane
direction due to the in-plane pressure asymmetry being normal to the
reconnecting magnetic field, are artificially suppressed in 2D.  It is
not expected drift waves will change the bulk properties of the
reconnection, but this is worth future study.

Another limitation of the present study is that it does not take into
account an out-of-plane component of bulk flow velocity or an
out-of-plane (guide) magnetic field.  It is known that diamagnetic
effects arise in systems with a guide field and a pressure asymmetry
\cite{Swisdak03}, and that the effects of flow shear and diamagnetic
drifts compound \cite{Tanaka10}.  The effect of out-of-plane flow has
been studied \cite{Wang12b,Ma14,Ma14b,Wang15}.  Incorporating
diamagnetic effects into the theoretical predictions
\cite{Pueschel15,Liu16} should be the subject of future work.

\begin{acknowledgments}
  Support from West Virginia University's Summer Undergraduate
  Research Experience (SURE) program (CED), the NASA West Virginia
  Space Grant Consortium (CED), NSF Grant AGS-0953463 (PAC), and NASA
  Grants NNX16AF75G and NNX16AG76G (PAC) are gratefully acknowledged.
  This research uses resources of the National Energy Research
  Scientific Computing Center (NERSC), a DOE Office of Science User
  Facility supported by the Office of Science of the U.S.~Department
  of Energy under Contract No.~DE-AC02-05CH11231.  We thank K.~Malakit
  for use of his color table in some of the plots, and we thank
  M.~T.~Beidler, C.~M.~Komar, and Y.~H.~Liu for stimulating
  discussions.  The simulation data used to produce the results of
  this paper are available from the authors by request.
\end{acknowledgments}


%

\end{document}